\documentclass[12pt]{iopart}
\usepackage{color,graphicx}
\usepackage{hyperref}

\begin{document}

\title{Effect of sine-Gaussian glitches on searches for binary coalescence}
\author{T Dal Canton$^1$, S Bhagwat$^2$, S V Dhurandhar$^3$ and A Lundgren$^1$}
\address{$^1$ Albert-Einstein-Institut, Max-Planck-Institut f\"ur Gravitationsphysik,
    D-30167 Hannover, Germany}
\address{$^2$ Indian Institute of Science Education \& Research, Central Tower,
    Sai Trinity Building, Pashan, Pune 411021, India}
\address{$^3$ Inter University Centre for Astronomy \& Astrophysics, Ganeshkhind,
    Pune - 411 007, India}
\ead{tito.dalcanton@aei.mpg.de}

\newcommand{\Sn}{S_{\rm n}}

\begin{abstract}
We investigate the effect of an important class of glitches occurring in the
detector data on matched filter searches of gravitational waves from coalescing
compact binaries in the advanced detector era. The glitches, which can be modeled
as sine-Gaussians, can produce triggers with significant time delays and thus
have important bearing on veto procedures as will be described in the paper.
We provide approximated analytical estimates of the trigger SNR
and time as a function of the parameters describing the sine-Gaussian (center
time, center frequency and $Q$-factor) and the inspiral filter (chirp mass).
We validate our analytical predictions through simple numerical simulations,
performed by filtering noiseless sine-Gaussians with the inspiral matched
filter and recovering the time and value of the maximum of the resulting SNR
time series. Although we identify regions of the parameter space in which each
approximation no longer reproduces the numerical results, the approximations
complement each other and together effectively cover the whole parameter space.
\end{abstract}


\section{Introduction}

Advanced interferometric gravitational-wave detectors are currently under
construction and are expected to start delivering strain data in a few years.
Major improvements over the first-generation interferometers will include
a significantly better sensitivity curve at all frequencies of interest, which
will effectively shift the sensitive band lower boundary from $\sim40$ Hz to
$\sim10$ Hz \cite{aligo, avirgo}.

A promising source of gravitational-wave detections in the advanced interferometer
era is represented by the inspiral and coalescence of compact binary systems,
such as binary neutron stars (BNS), binary black holes and black hole-neutron star binaries
\cite{cbcrates}. The strain waveform associated with such events can be calculated
to a good approximation at least for some regions of the parameter space describing
the binary system \cite{blanchetlrr}. Consequently, the classical data analysis
pipeline for detecting such signals employs a filter matched to the analytical
waveform \cite{ihope}. This approach consists in correlating the strain data with
the waveform and producing an estimate of the signal to noise ratio (SNR) for
that waveform. Triggers are then generated whenever the SNR exceeds a pre-defined
threshold. Lists of triggers are independently produced for each detector and
combined to obtain coincident triggers, which are then checked for consistency
and passed on to more elaborate analysis. Similar
search pipelines will be employed in the advanced detector era. However,
in order to exploit the improved noise curves and extend the search to lower
frequencies, the strain data will have to be filtered with significantly longer
inspiral waveforms than used for past searches (up to tens of minutes as opposed
to about one minute for BNS). In fact, the duration of the
waveform in the interferometer output depends strongly on the low-frequency
cutoff \cite{findchirp}.

First generation interferometers were affected by high rates of short-duration
transients unrelated to astrophysical events and commonly known as \emph{glitches},
most of which can be roughly modeled as short sine-Gaussian waveforms with a few
cycles, i.e.~a small $Q$-factor \cite{lscglitchgroup}.
The typical effect of glitches on searches for binary coalescence is a
significant deviation of the SNR distribution from what expected for stationary
Gaussian noise, i.e.~an exponential distribution. In particular,
glitches produce tails of high-SNR outliers unrelated to astrophysical signals.
This increases the false-alarm rate of the search and lowers our confidence on
possible detections of true inspiral signals (see \cite{virgodetchar}, figure 8).
In past searches, most glitches were ruled out as gravitational-wave candidates
by monitoring \emph{auxiliary channels} and their origin could rather be
attributed either to the noisy environment of the detector (for instance
seismic noise, weather, electromagnetic disturbances) or to unexpected behavior
of the detector itself (saturation of analog-to-digital converters, problems with control systems and
thermal compensation, scattered light) \cite{s5glitches}. Nevertheless, surviving
glitches still increased the rate of coincident triggers between detectors and
thus weakened our ability to both confidently detect signals as well as set
strong upper limits on the coalescence rate.
Unfortunately, despite the improved sensitivity, advanced detectors are expected
to manifest similar glitches in their output. This will still pose significant
challenges to future gravitational-wave searches and could in fact be their
ultimate limiting factor. Hence, studies investigating the effect of data
quality on searches play a central role in future results.

A potential complication of advanced inspiral searches is the fact that the
longer BNS inspiral filters could lead to significant time delays between the
occurrence of a short, isolated glitch and the coalescence time of the resulting
spurious inspiral trigger. Often, such delays will be much longer than the
glitch itself and they may have an impact both on existing veto procedures based
on auxiliary channels and on the tuning of consistency checks such as Allen's
$\chi^2$ \cite{chisquared}, bank $\chi^2$ and auto $\chi^2$ \cite{chisquared2}.
Being able to predict the parameters of spurious inspiral triggers produced by
glitches is important for understanding such issues and could hint at new veto
techniques for reducing false-alarms produced by glitches.

For this reason, we study the effect of glitches with sine-Gaussian waveform on
a matched-filter search for inspiral gravitational waves in the advanced
detector era. We present approximations to the complex matched-filter output
and to the SNR time series (the quantity used for producing the inspiral
triggers) for the case where the signal is just a noiseless sine-Gaussian with
no additional signal components on top of it. Given the parameters of the glitch
model (central time $t_0$, central frequency $f_0$ and quality factor $Q$) such
approximations allow one to predict the SNR and time of the resulting spurious
triggers as a function of the chirp mass $\mathcal{M}$ of each template. We
present three such approximations which complement each other and completely
cover the parameter space of the sine-Gaussians. We support the approximations
with numerical simulations.

The paper is organized as follows. In section 2 we briefly review the matched
filter algorithm and fix the relevant notation. In section 3 we derive the
approximated estimates for the SNR time series and trigger parameters. In
section 4 we compare these results to numerical simulations and we identify
regimes where the different approximations are no longer acceptable. Section 5
discusses the implications of our results and suggests a way to use such
results for constructing a veto procedure for the advanced era.

\section{Matched filter}

A matched filter search combines a detector strain data segment $s$, the signal
to be sought $h$ and the one-sided power spectral density (PSD) of the
detector noise $\Sn(f)$ into a quantity representing the SNR as a function
of the parameters describing the gravitational-wave source \cite{findchirp}.
The SNR for a binary coalescing at time $t_{\rm c}$ is calculated as
\footnote{For simplicity and since search algorithms work with SNR time series,
we only explicitly display the dependency on the coalescence time $t_{\rm c}$ and omit
other parameters.}
\begin{equation}
	\rho(t_{\rm c}) = \frac{|z(t_{\rm c})|}{\sigma}
	\label{rho}
\end{equation}
where $z(t_{\rm c})$ is the output of the complex matched filter,
\begin{equation}
	z(t_{\rm c}) = 4 \int_{f_{\rm L}}^{f_{\rm H}} \frac{\tilde{s}(f) \tilde{h}^*(f;t_{\rm c})}{\Sn(f)} \rmd f
	\label{mfo}
\end{equation}
and $\sigma$ is the sensitivity of the detector to the sought after signal,
\begin{equation}
	\sigma^2 = 4  \int_{f_{\rm L}}^{f_{\rm H}} \frac{|\tilde{h}(f;t_{\rm c})|^2}{\Sn(f)} \rmd f.
	\label{sigmasq}
\end{equation}
In the above expressions, $\tilde{s}(f)$ is the Fourier transform of the
pre-conditioned data segment and
\begin{equation}
	\tilde{h}(f;t_{\rm c}) = h_0 f^{-7/6} e^{-i \psi(f;t_{\rm c})}
	\label{template}
\end{equation}
is the inspiral signal template expressed in the frequency domain via the
stationary phase approximation \cite{cbcspa}. $f_{\rm L}$ and $f_{\rm H}$ are frequency cutoff values.
Searches performed on first-generation detector data used $f_{\rm L} = 40$ Hz and for
advanced interferometers $f_{\rm L} = 10$ Hz is expected. $f_{\rm H}$ is usually set to the
frequency corresponding to the innermost stable circular orbit of a test mass
orbiting a Schwarzschild black hole, i.e.~$f_{\rm ISCO} = c^3 / (6 \sqrt{6} \pi G M)$.

Candidate inspiral triggers are identified as local maxima of the SNR
crossing a pre-established threshold. Each trigger carries the set of
associated parameter values, e.g.~coalescence time and phase and binary
component masses.

\section{Response to a sine-Gaussian waveform}

We are interested in calculating the response of the inspiral matched filter
to a sine-Gaussian waveform, i.e.~the SNR time series $\rho(t_{\rm c})$ after the center
time of the sine-Gaussian. From $\rho(t_{\rm c})$ we then want to estimate the maximum
SNR $\hat{\rho}$ and the corresponding time $\hat{t}_{\rm c}$. The signal we consider
includes only a sine-Gaussian glitch: neither the stochastic noise of the
detector nor any kind of astrophysical gravitational wave signals are added.
Therefore, our results will represent averages over the ensemble of noise
realizations. They will also not describe the case of a glitch contaminating
a true inspiral signal.

The sine-Gaussian waveform can be written in the time domain as
\begin{equation}
	s(t) = A \exp \left( -\frac{(t-t_0)^2}{\tau^2} \right) \cos (2\pi f_0 t + \phi_0)
\end{equation}
where $A$ is an overall amplitude, $t_0$ and $f_0$ are the location of the
sine-Gaussian in time and frequency respectively, $\tau = Q / 2\pi f_0$ is the
time duration, $Q$ is the dimensionless quality factor and $\phi_0$ is the phase
of the oscillation at $t = 0$. We are free to define the time origin as $t_0 = 0$,
so that $\hat{t}$ represents the time delay between the sine-Gaussian and the
inspiral trigger. Since the SNR is linear in the signal amplitude, we can also
set $A = 1$. The sine-Gaussian can then be expressed in the frequency domain as
\begin{equation}
	\tilde{s}(f) = \frac{\sqrt{\pi}}{2} \tau \exp \left( -\pi^2 \tau^2
		(f-f_0)^2 + i\phi_0 \right) \left[ 1 + \exp \left( -Q^2 \frac{f}{f_0}
        - 2i\phi_0 \right) \right].
	\label{sgfreqdom}
\end{equation}
For $f_{\rm L} < f < f_{\rm H}$, the second exponential falls off very quickly as
$Q$ increases from 1, so we neglect it altogether. We can now also set $\phi_0 = 0$
since the SNR is not affected by global phase factors. Inserting \eref{template}
and the so-modified \eref{sgfreqdom} into \eref{mfo} we get
\begin{equation}
	z(t_{\rm c}) = 2 \sqrt{\pi} \tau h_0
		\int_{f_{\rm L}}^{f_{\rm H}} \frac{f^{-7/6}}{\Sn(f)} \exp \left(
		-\pi^2 \tau^2 (f - f_0)^2 + i \psi(f;t_{\rm c}) \right) \rmd f.
	\label{sgmfo}
\end{equation}
In order to get an explicit expression, we approximate $\psi(f;t_{\rm c})$ with the simple
\emph{Newtonian chirp} \cite{petersmathews}
\begin{equation}
	\psi(f;t_{\rm c}) = 2\pi f t_{\rm c} - 2\phi -\frac{\pi}{4} + \frac{3}{128} (\zeta f)^{-5/3}
\end{equation}
where $t_{\rm c}$ is the coalescence time, $\phi$ is the orbital phase at coalescence, $\zeta = \pi G
\mathcal{M} / c^3$ and $\mathcal{M} = (m_1 m_2)^{3/5}/(m_1 + m_2)^{1/5}$ is the
chirp mass of the binary system composed of masses $m_1$ and $m_2$. Although the
resulting integral can not be evaluated exactly, different approximations can be
found, as will be discussed in the next subsections.

\subsection{Approximation I}

The first approximation we describe makes use of the stationary phase
approximation. If $\rmd\psi/\rmd f = 0$ at a frequency $f_{\rm s}$, we can
approximate $\psi(f;t_{\rm c})$ around $f_{\rm s}$ with a second order power series,
\begin{equation}
	\psi(f;t_{\rm c}) \approx \psi(f_{\rm s}) + \beta (f - f_{\rm s})^2
	\label{psispa}
\end{equation}
For the Newtonian chirp this holds and we have
\begin{eqnarray}
	f_{\rm s} &=& \left( \frac{5}{256 \pi} \right)^{3/8} \zeta^{-5/8} t_{\rm c}^{-3/8}
		= f_0 \left(\frac{\tau_0}{t_{\rm c}}\right)^{3/8} \\
	\beta &=& \frac{5}{96} f_{\rm s}^{-11/3} \zeta^{-5/3}
\end{eqnarray}
where $\tau_0$ is the first \emph{chirp time} evaluated using $f_0$ as the
fiducial frequency \cite{chirptimes}. Note that $f_{\rm s}$ is a function of the
coalescence time and we are evaluating a family of integrals parametrized by $t_{\rm c}$.
The factor $f^{-7/6} \Sn^{-1}(f)$ in \eref{sgmfo} varies more slowly with
frequency than the exponential, so we take it as constant and evaluate it at
$f_{\rm s}$. The integrand can then be rewritten as the product of two Gaussians, one
real and one complex, so the matched filter output is (omitting irrelevant
phase factors)
\begin{equation}
	z(t_{\rm c}) \approx 2\sqrt{\pi}\tau \frac{f_{\rm s}^{-7/6}}{\Sn(f_{\rm s})} \int_{f_{\rm L}}^{f_{\rm H}} \exp \left(
		i \frac{(f - f_{\rm s})^2}{2\sigma^2_{\rm f}} - \frac{(f - f_0)^2}{2\sigma^2_{\rm sg}}\right) \rmd f
	\label{mfo1}
\end{equation}
where $\sigma_{\rm sg} = 1/\sqrt{2} \pi \tau$ and $\sigma_{\rm f} = \sqrt{3 f_0 / 16 \pi \tau_0}
(f_{\rm s} / f_0)^{11/6}$ are the standard deviations of the Gaussians. Using
\eref{mfo1} and \eref{sigmasq} in \eref{rho}, and carrying out the integral
using Cauchy's theorem, we get
\begin{equation}
	\rho(t_{\rm c}) \approx \frac{1}{\sigma_0} \frac{\exp \left(
		-\frac{(f_0 - f_{\rm s})^2}{2\sigma^2_{\rm f}} \frac{\sigma_{\rm sg}^2 / \sigma_{\rm f}^2}{1+
		\sigma_{\rm sg}^4 / \sigma_{\rm f}^4} \right)}
		{f_{\rm s}^{7/6} \Sn(f_{\rm s}) (1 + \sigma_{\rm sg}^4 / \sigma_{\rm f}^4)^{1/4}}
\end{equation}
with $\sigma_0^2 = \int_0^{\infty} f^{-7/3} \Sn^{-1}(f) \rmd f$.

This is a complicated function of $t_{\rm c}$ through $f_{\rm s}$ and $\sigma_{\rm f}$. In particular
note that $t_{\rm c}$ also enters via $\Sn(f_{\rm s})$ and so the shape of the SNR pulse
in time depends on the PSD of the noise. Although $\Sn(f)$ can in principle be modeled
as a combination of power laws, the resulting expression is complicated and a
numerical maximization is thus needed in order to determine $\hat{\rho}$ and $\hat{t}_{\rm c}$.

\subsection{Approximation II}

Our second method is a slightly different application of the stationary phase
approximation and is expected to be appropriate for small $Q$. In fact, we still
use \eref{psispa} but we now assume that the largest contribution to the
integral in \eref{sgmfo} comes from a narrow band around $f_{\rm s}$.
Within this band, we can thus consider all other terms of the integrand
constant and equal to their value at $f_{\rm s}$, including the sine-Gaussian peak,
so the integral reduces to
\begin{equation}
	\int_{f_{\rm L}}^{f_{\rm H}} e^{i \beta (f - f_{\rm s}) ^2} \rmd f \approx \sqrt{\frac{\pi}{\beta}}.
\end{equation}
Therefore, omitting all irrelevant phase terms,
\begin{equation}
	z(t_{\rm c}) \approx 2 \pi \tau h_0 \frac{\exp \left(-\pi^2 \tau^2 
		(f_{\rm s} - f_0)^2\right)}{f_{\rm s}^{7/6} \Sn(f_{\rm s}) \beta^{1/2}}
\end{equation}
and the resulting SNR time series is
\begin{equation}
	\rho(t_{\rm c}) \approx \frac{\pi \tau}{\sigma_0} \frac{\exp \left(-\pi^2 \tau^2
		(f_{\rm s} - f_0)^2\right)}{f_{\rm s}^{7/6} \Sn(f_{\rm s}) \beta^{1/2}}.
\end{equation}
For the Newtonian chirp,
\begin{equation}
    \rho(t_{\rm c}) \approx \frac{\sqrt{6} \pi^{3/4}}{5^{1/4}}
        \frac{\tau \zeta^{5/12}}{\sigma_0}
        \frac{\exp \left(-\pi^2 \tau^2 (f_{\rm s} - f_0)^2\right)}{\Sn(f_{\rm s})} t_{\rm c}^{-1/4}.
\end{equation}
This is still a quite complicated function of time and it still depends on the
PSD of the noise. In general, at least for a smooth PSD with no peaks, it is
the product of two peaks with different
time scales, one corresponding to the exponential term and the other associated
with the $f_{\rm s}^{7/6} \Sn(f_{\rm s}) \beta^{1/2}$ term. Thus $\hat{t}_{\rm c}$ depends on $f_0$,
$\tau$, $\zeta$ and the details of $\Sn(f)$ even for the Newtonian chirp.
Although we expect approximation II to be valid for small $Q$, we can try to
take the result to the limit of $Q\to \infty$. The exponential becomes more peaked
than the $f_{\rm s}^{-7/6} \Sn^{-1}(f_{\rm s}) \beta^{-1/2}$ term, so $\rho(t_{\rm c})$ is maximum
where $f_{\rm s} = f_0$ and
\begin{eqnarray}
	\hat{\rho} &\approx \frac{\pi \tau}{\sigma_0} f_0^{-7/6} \Sn^{-1}(f_0) \beta^{-1/2} \\
	\hat{t}_{\rm c} &\approx \tau_0.
\end{eqnarray}
This suggests that in the large $Q$ limit the trigger delay is just the time the
inspiral takes to coalesce after crossing the center frequency of the
sine-Gaussian, which can be expected from intuition. As we said, however, this
approximation is expected to work for small $Q$ and thus in general $\hat{\rho}$
and $\hat{t}_{\rm c}$ must be found numerically.

\subsection{Approximation III}

Although probably less useful in practice, a different approximation can be
derived in the large $Q$ limit. In this regime the integrand in \eref{sgmfo}
is no longer dominated by the
band around $f_{\rm s}$ but by the narrow peak of the sine-Gaussian, centered on $f_0$.
Again, the factor $f^{-7/6} \Sn^{-1}(f)$ varies slowly with frequency and can be
regarded as constant, and the integral limits can be extended over the real axis.
We still approximate the template phase as a second-order Taylor series, but
this time around $f_0$,
\begin{equation}
	\psi(f;t_{\rm c}) \approx \psi(f_0) + \alpha (f - f_0) + \beta (f - f_0)^2.
\end{equation}
For instance, for the Newtonian chirp we have
\begin{eqnarray}
	\alpha &=& 2\pi t_{\rm c} - \frac{5}{128} f_0^{-8/3} \zeta^{-5/3} = 2\pi(t_{\rm c} - \tau_0) \\
	\beta &=& \frac{5}{96} f_0^{-11/3} \zeta^{-5/3}.
\end{eqnarray}
After switching the integration variable to $x = \sqrt{\pi^2 \tau^2 -
i \beta } (f - f_0)$, and omitting irrelevant phase factors, we are left with a Gaussian integral
\begin{equation}
	z(t_{\rm c}) \approx \frac{2 \sqrt{\pi} \tau h_0}{\sqrt{\pi^2 \tau^2 - i \beta}}
		\frac{f_0^{-7/6}}{\Sn(f_0)}
		\int_{-\infty}^{+\infty} \exp \left( -x^2
		+ \frac{i \alpha x}{\sqrt{\pi^2 \tau^2 - i \beta}} \right) \rmd x
\end{equation}
which readily gives
\begin{equation}
	z(t_{\rm c}) \approx \frac{2 \pi \tau h_0}{\sqrt{\pi^2 \tau^2 - i \beta}}
		\frac{f_0^{-7/6}}{\Sn(f_0)}
		\exp \left( -\frac{\alpha^2}{4(\pi^2 \tau^2 - i \beta)} \right).
\end{equation}
The final expression for the approximated SNR time series is
\begin{equation}
	\rho(t_{\rm c}) \approx \frac{1}{\sigma_0} 
		\frac{\exp \left( -\alpha^2 / 4 \pi^2 \tau^2 (1+\frac{\beta^2}{\pi^4\tau^4})
		\right)}{f_0^{7/6} \Sn(f_0) \left( 1 + \frac{\beta^2}{\pi^4 \tau^4}\right)^{1/4}}.
\end{equation}
In the case of the Newtonian chirp $\alpha$ is the only function of $t_{\rm c}$,
so $\rho(t_{\rm c})$ is a simple Gaussian pulse and we obtain explicit expressions for
trigger SNR and time,
\begin{eqnarray}
	\hat{\rho} &\approx \sigma^{-1}_0 f_0^{-7/6} \Sn^{-1}(f_0) \left( 1 + \frac{\beta^2}{\pi^4\tau^4}\right)^{-1/4} \\
	\hat{t}_{\rm c} &\approx \tau_0.
\end{eqnarray}
Note that $\hat{t}_{\rm c}$ is consistent with approximation II taken in the limit of
large $Q$, but $\hat{\rho}$ is not and we expect approximation III to give the
correct result in this case.

\section{Numerical simulations}

\begin{figure}
\begin{center}
\includegraphics[width=\textwidth]{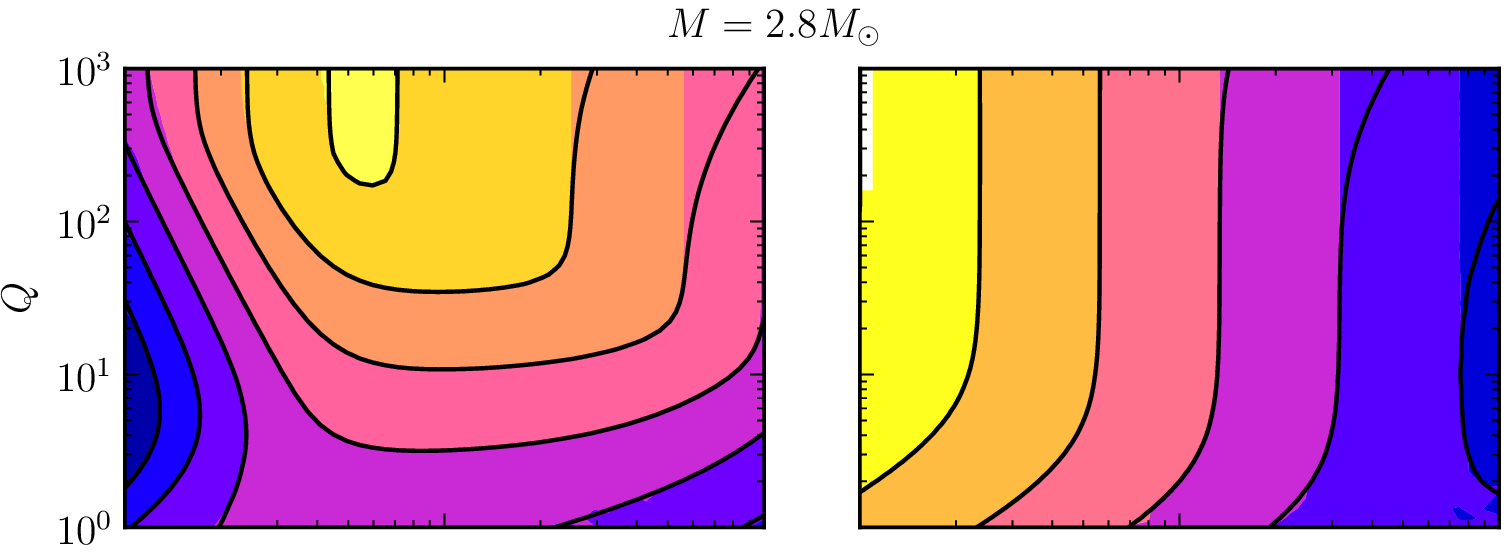}
\includegraphics[width=\textwidth]{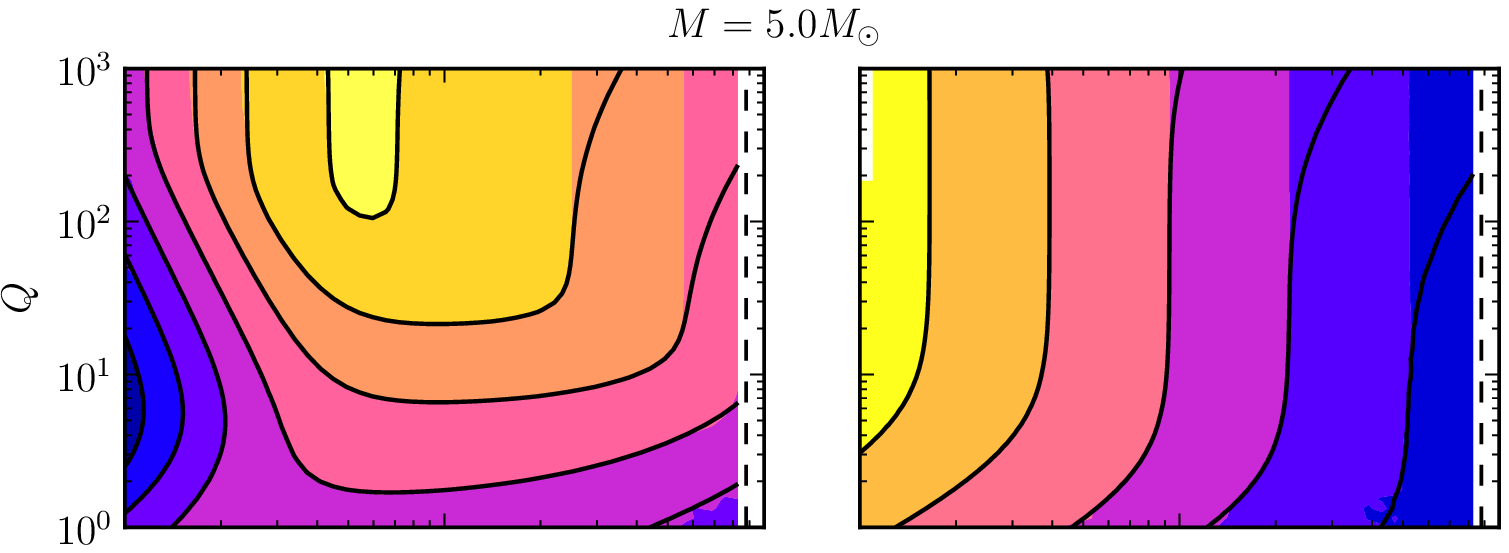}
\includegraphics[width=\textwidth]{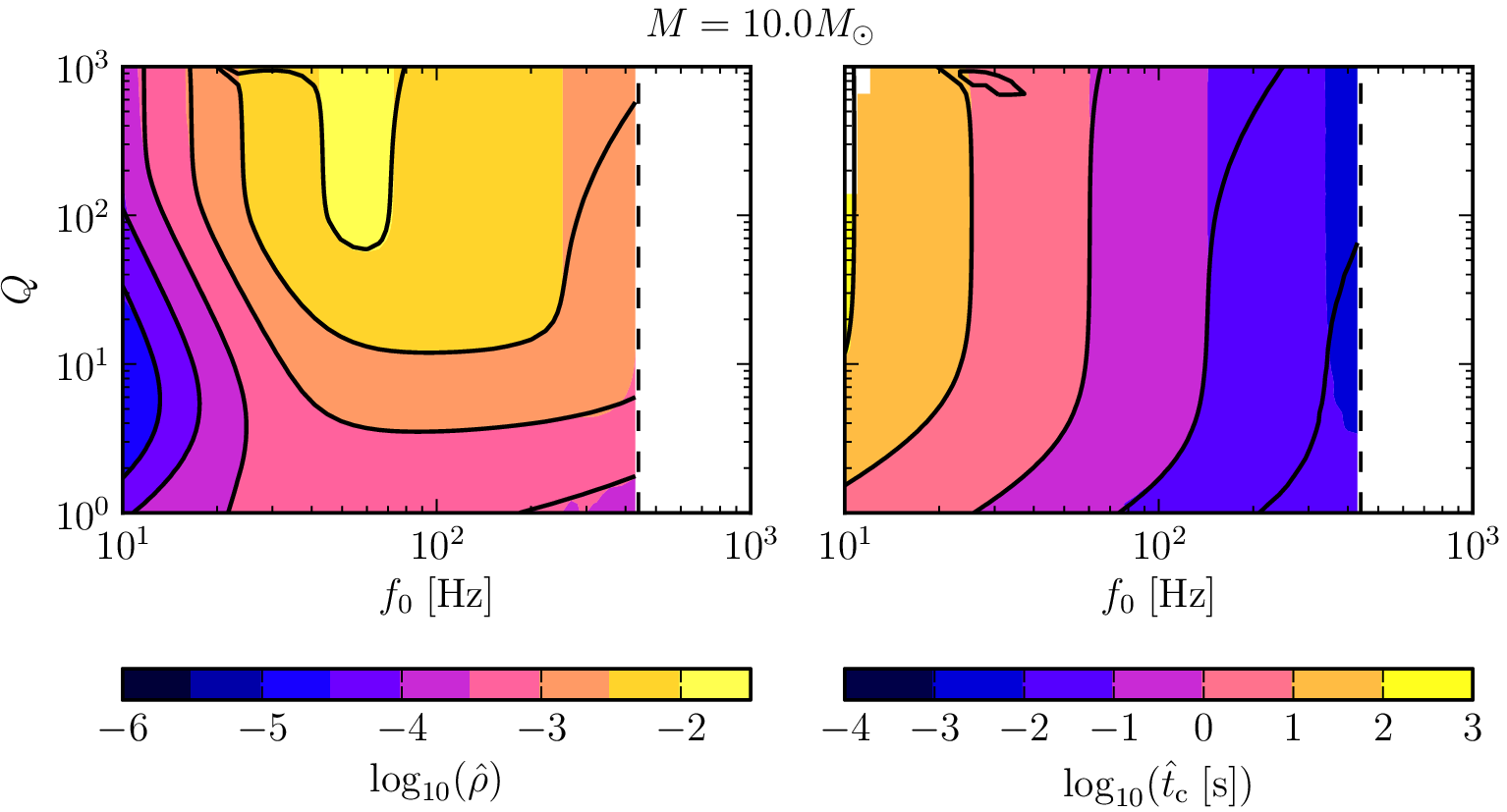}
\end{center}
\caption{Comparison between simulations (shaded bands, color online) and
approximation I (black contours) for different total masses. Left plots show
the trigger SNR, right ones the delay. The dashed line is the ISCO frequency.
As can be seen, this approximation fails for large $f_0$, $Q$ or $M$.}
\label{contours_approx_I}
\end{figure}

\begin{figure}
\begin{center}
\includegraphics[width=\textwidth]{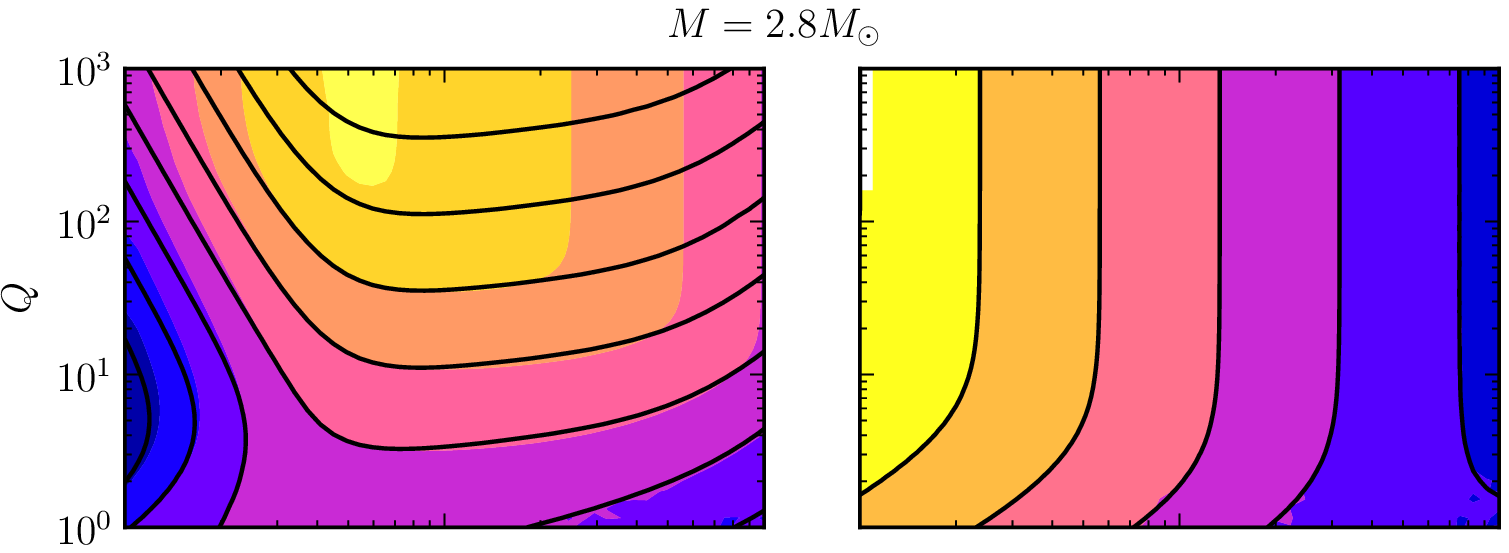}
\includegraphics[width=\textwidth]{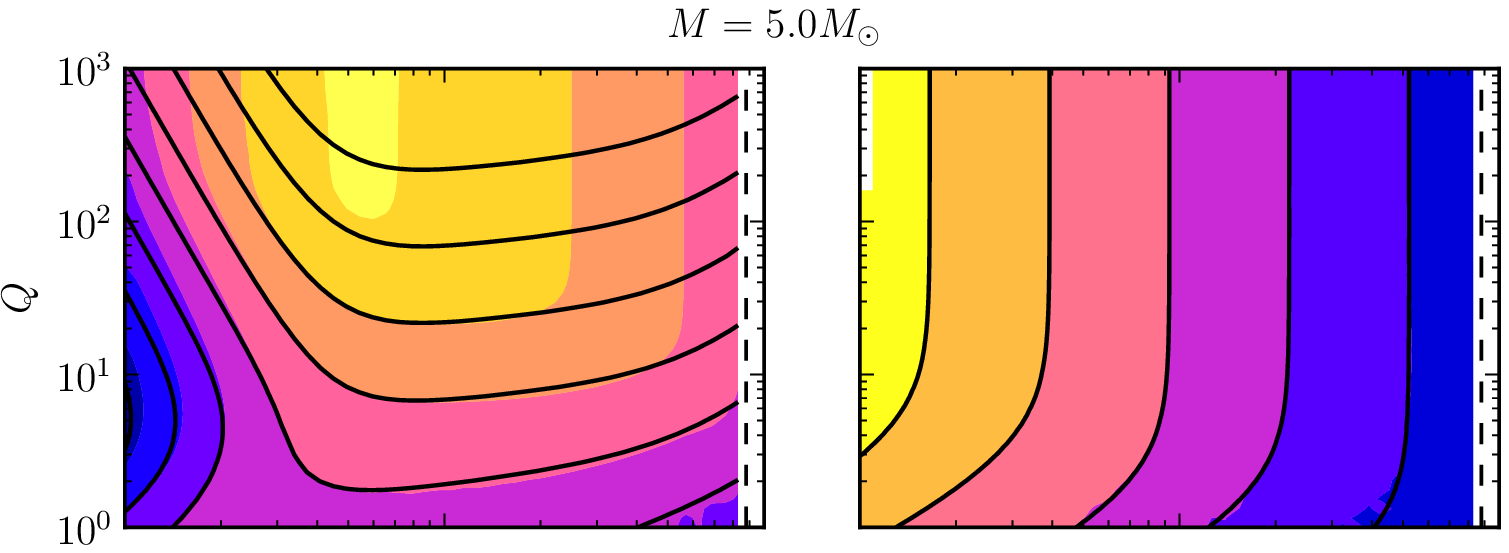}
\includegraphics[width=\textwidth]{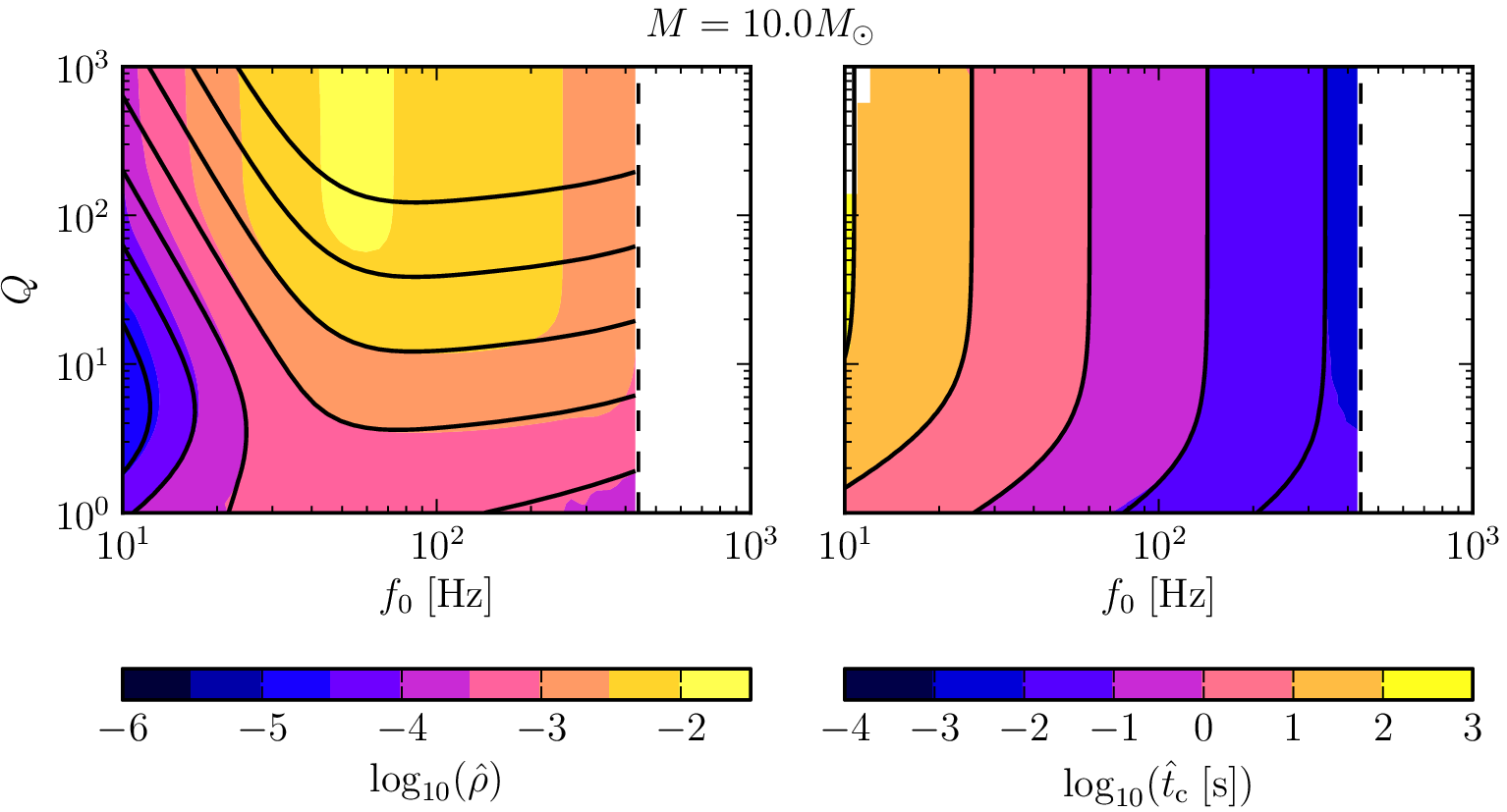}
\end{center}
\caption{Comparison between simulations (shaded bands, color online) and
approximation II (black contours) for different total masses. Left plots show
the trigger SNR, right ones the delay. The dashed line is the ISCO frequency.
The SNR predicted by approximation II is not correct for $Q \gg 1$, as
expected. Remarkably however, the predicted trigger time is good for most of
the explored parameter space.}
\label{contours_approx_II}
\end{figure}

\begin{figure}
\begin{center}
\includegraphics[width=\textwidth]{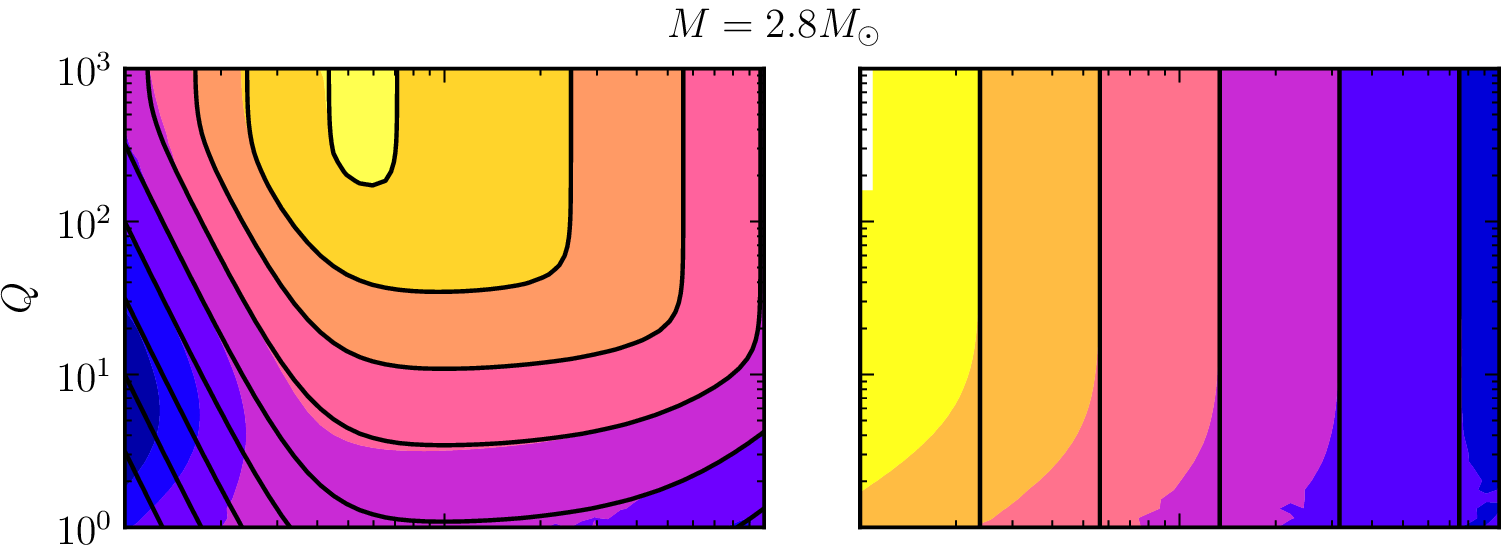}
\includegraphics[width=\textwidth]{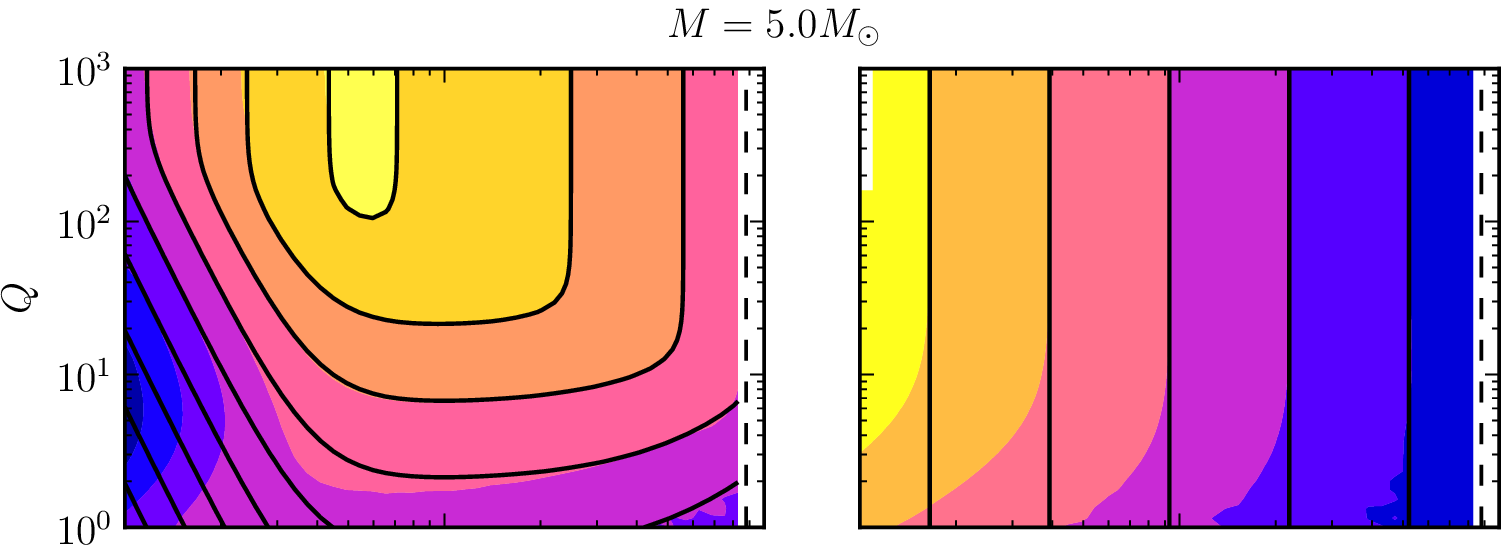}
\includegraphics[width=\textwidth]{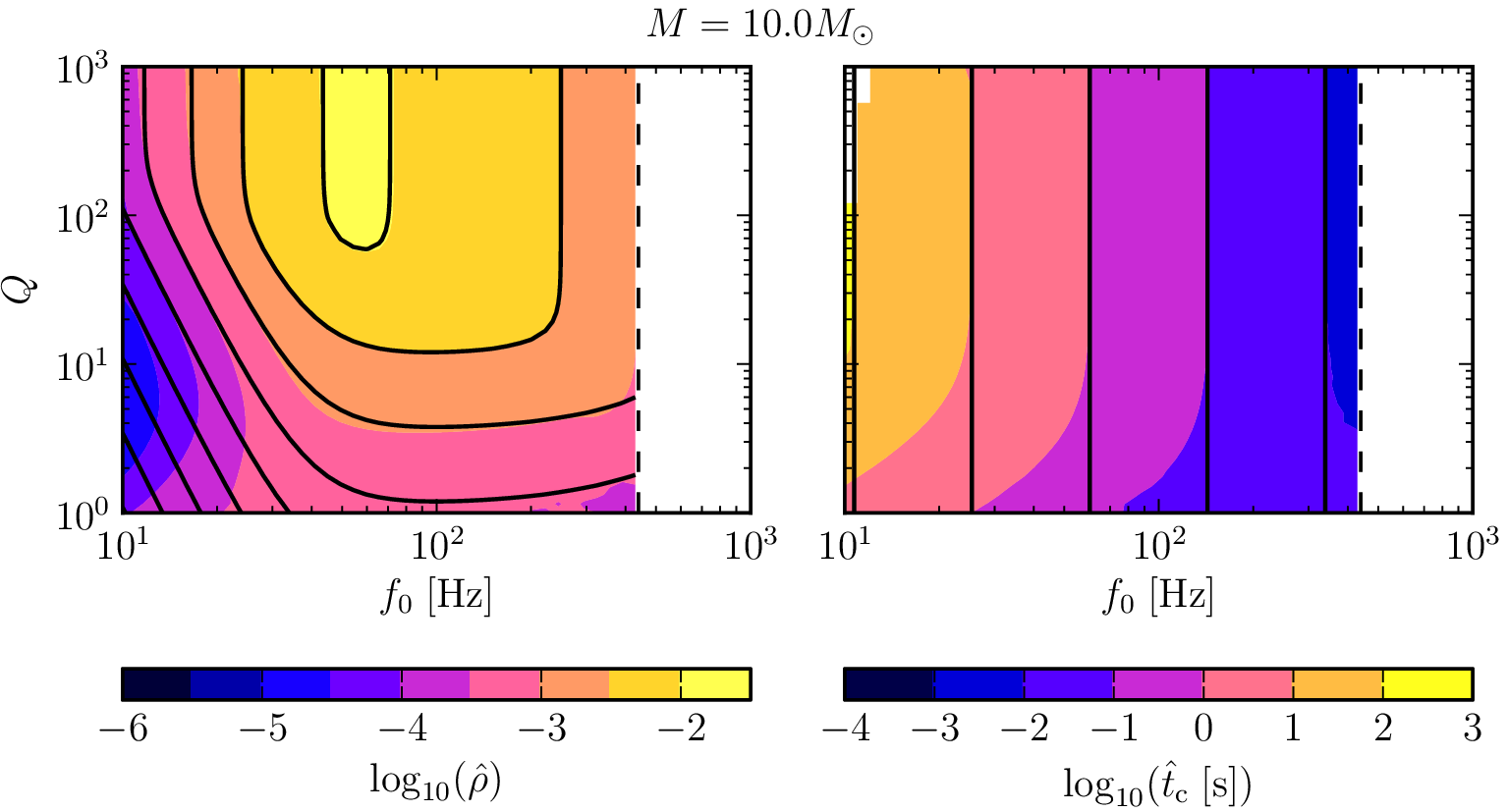}
\end{center}
\caption{Comparison between simulations (shaded bands, color online) and
approximation III (black contours) for different total masses. Left plots show
the trigger SNR, right ones the delay. The dashed line is the ISCO frequency.
As expected, approximation III does not work for small $Q$.}
\label{contours_approx_III}
\end{figure}

We test the accuracy of the approximations by numerically computing the response
of an inspiral matched filter to noiseless sine-Gaussians injected in the time domain.
For $\Sn(f)$ we take the noise PSD corresponding to advanced LIGO's \emph{zero
detuning, high power} configuration, an approximation of which is provided by
the LALSimulation module of LALSuite \cite{lalsimulation}. For simplicity, and in order to
work well within the numerical range of the floating-point representation, we
set the amplitude of the simulated sine-Gaussian to 1 and at the same time we
scale the noise PSD by $10^{48}$; therefore, the resulting trigger SNR refers
to sine-Gaussians with an amplitude of $A = 10^{-24}$.
We scan the $(f_0, Q)$ parameter space in the region $10 \textrm{ Hz} \le f_0 \le
\min \left(1 \textrm{ kHz}, f_{\rm ISCO}\right)$, $1 \le Q \le 1000$ for total mass values of $2.8 M_\odot$,
$5 M_\odot$ and $10 M_\odot$. For each point, we compute the maximum SNR
$\hat{\rho}$ and the delay $\hat{t}_{\rm c}$ between the maximum SNR and the center
of the injected sine-Gaussian, and we compare to the predicted values. Results
for approximation I, II and III can be seen in figure  \ref{contours_approx_I},
\ref{contours_approx_II} and \ref{contours_approx_III} respectively.

Approximation I works well for both $\hat{\rho}$ and $\hat{t}_{\rm c}$, but it degrades
for large values of $f_0$, $Q$ or mass. In particular, we find that surfaces
of constant accuracy roughly match those of constant $f^2_0 Q M$; the 5\%
accuracy for SNR, at least in the explored parameter range, is at $f^2_0 Q M /
M_\odot \approx 5 \times 10^7$ s$^{-2}$. This is likely not a major problem, as
we expect most glitches to last at most tens of cycles and affect mostly low
frequencies. Moreover, at high frequency or mass the time delay is small and
thus the problem we are considering is less important. Also note that neglecting
the second exponential of \eref{sgfreqdom} produces no noticeable effects
even for $Q \approx 1$.

Approximation II produces excellent estimates of the time delay across all
the explored parameter space, but wrong estimates of SNR for large $Q$, as can
be expected. In fact, as $Q$ increases, the sine-Gaussian peak in the frequency
domain shrinks and at some point the integral is no longer dominated by the
region around $f_{\rm s}$. Since we are interested in the low-$Q$ region, despite
this problem this is still a useful approximation, although not significantly
simpler than I.

As expected, approximation III works very well in the high mass, high $Q$ region
where the other two approximations do not give such good results. Moreover
it is analytically simpler. But for a large region of the parameter space where
$Q$ is small, this approximation fails and so is not so useful for a large part
of the parameter space of interest. This is because the width of the Gaussian
in the frequency domain becomes comparable or larger than the width of the other
terms in the integrand and the approximation is not valid. However, in this part
of the parameter space we can simply revert to approximations I and II.

\begin{figure}
\begin{center}
\includegraphics[width=\textwidth]{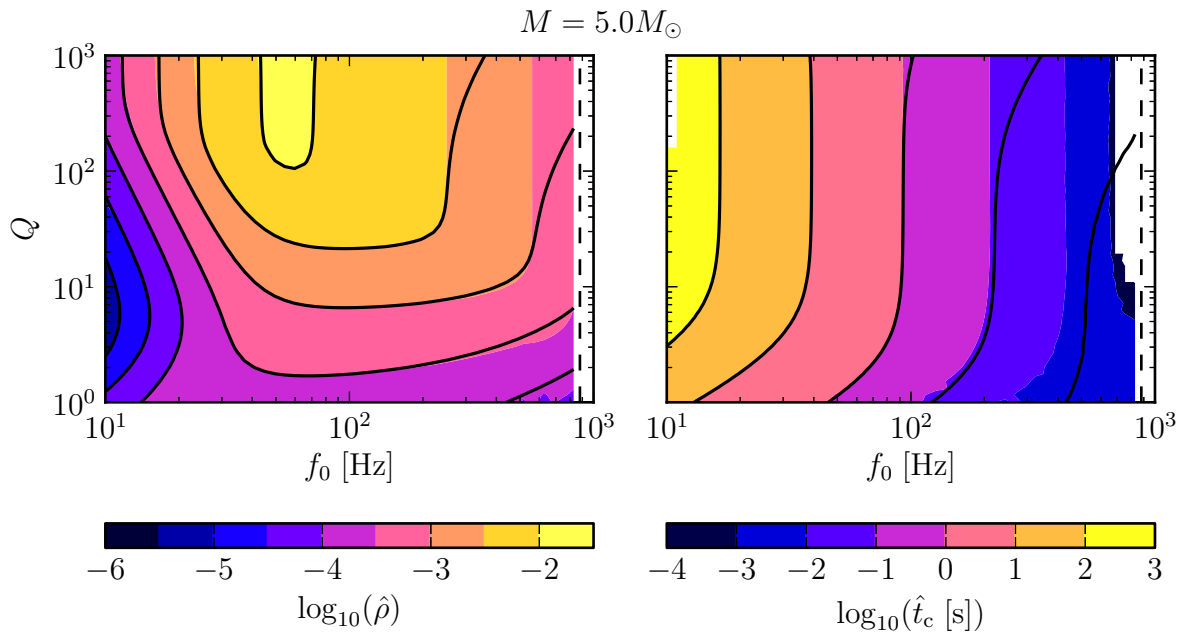}
\end{center}
\caption{Comparison between a simulation using a 3.5PN equal-mass inspiral
waveform (shaded bands, color online) and approximation I (black contours) for
a total mass of $5 M_\odot$. The left plot is the trigger SNR, the right one
is the trigger delay. The dashed line is the ISCO frequency.}
\label{contours_approx_I_3pn}
\end{figure}

Although our calculations are based on the Newtonian chirp, search pipelines use
inspiral waveforms with higher post-Newtonian order \cite{findchirp}. Thus
we also check the accuracy of the approximations against simulations using a 3.5PN
inspiral filter. We find that the accuracy degrades but is still within a few
percent both for $\hat{\rho}$ and $\hat{t}_{\rm c}$ and it retains a similar dependency
on $f_0$, $Q$ and $M$. An example for a $5 M_\odot$ equal-mass binary is shown
in figure \ref{contours_approx_I_3pn}. Since our approximations are based simply
on power-series approximations of the inspiral phase, they can in principle
accommodate high post-Newtonian order waveforms, at the price of more
complicated expressions for $\rho(t_{\rm c})$.

\begin{figure}
\begin{center}
\includegraphics[width=\textwidth]{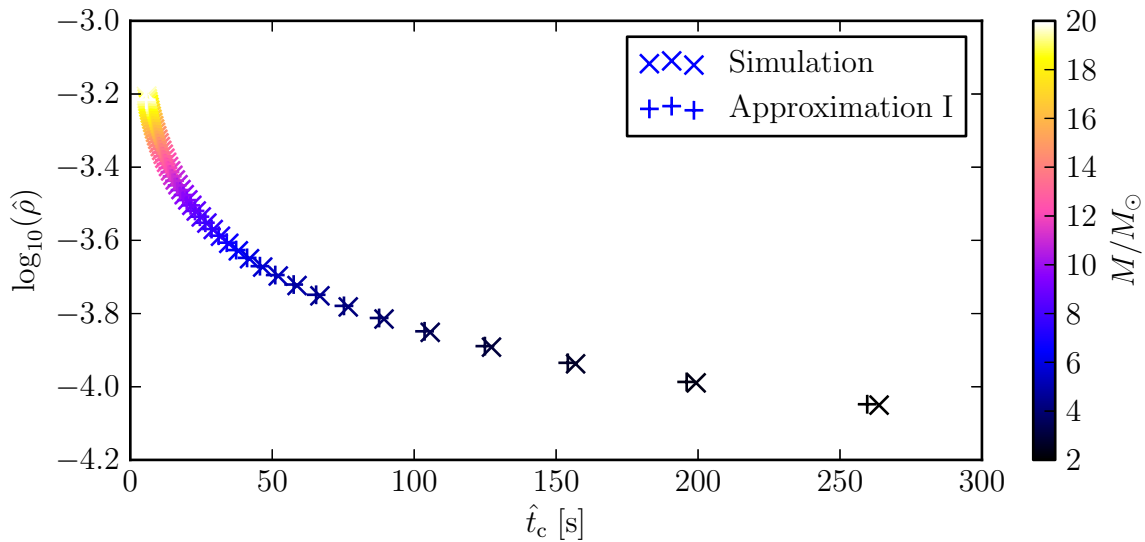}
\end{center}
\caption{Cluster of triggers generated by a template bank responding to a
sine-Gaussian glitch with $t_0=0$, $f_0=20$ Hz and $Q=20$. The bank consists
of 50 3.5PN, equal-mass waveforms with $M$ uniformly spaced in $[2 M_\odot,
20 M_\odot]$. The cluster extends well after the duration of the glitch,
with the lowest mass templates still producing triggers minutes after
the glitch.}
\label{cluster}
\end{figure}

In a search pipeline, the parameter space of the binary is covered
by a template bank. A true inspiral signal produces triggers only for
templates whose parameters are close enough to the true values, depending on
the ambiguity function of the waveform. However, a strong glitch generally
excites a significant fraction of the whole bank, producing a cluster of
triggers with different SNRs and times.
An example of this phenomenon is shown in figure \ref{cluster},
where we plot the distribution of triggers in $(\hat{t}_{\rm c}, \hat{\rho})$
corresponding to a sine-Gaussian glitch affecting a simplified template bank
with uniform distribution in total mass. The dependency of SNR and time delay
on the template mass determines the shape of the cluster. Low-mass templates
produce the triggers with the largest delay and smallest SNR. The last
triggers can be delayed by several minutes, which is much longer than the
duration of the glitch (typically seconds or fractions of a second).

Again we stress that, since we considered the response of the matched filter to
a noiseless sine-Gaussian glitch, the predicted coalescence time and SNR
represent expectation values. Triggers produced by glitches in real data will
have parameters distributed around our prediction. In particular we expect
that, in the limit of high SNR, such distributions will tend to Gaussians
centered on our prediction.

\section{Conclusion}

The length of low-mass inspiral filters represents a potential problem for
future inspiral searches and in particular for low-latency BNS pipelines such as
MBTA \cite{mbta} and LLOID \cite{lloid}. In fact, as evident from this study,
short glitches with sine-Gaussian-like waveforms can produce low-mass triggers
well after their occurrence time. Combined with the fact that a glitch can have
significant overlap with a large fraction of the template bank, this implies
that a single glitch can produce a cluster of many spurious triggers spanning a
time interval of several minutes. Although we found that the trigger SNR
effectively decreases with increasing delay, glitches can be very strong and
still produce triggers with a large delay and SNR above the detection threshold.
Consistency tests such as Allen's $\chi^2$ \cite{chisquared}, bank $\chi^2$ and
auto $\chi^2$ \cite{chisquared2} are most effective for long waveforms and would
likely rule out a large fraction of such spurious triggers. A detailed
investigation of the response and effectiveness of these tests with respect to
sine-Gaussian waveforms is a necessary followup of this work and could provide
hints at how to optimally tune the parameters of such tests for this particular
glitch model.

Unfortunately however, the efficiency of consistency tests also decreases with
decreasing SNR. Triggers with large delay and SNR just above the detection
threshold will therefore still be problematic and the usual veto procedures
based on auxiliary channels will be required. For past searches, such
procedures consisted in identifying glitches in one or more auxiliary channels
and excluding inspiral triggers within an appropriate coincidence window. While
this method is applicable in the case of short inspiral filters, though, its
naive application to future low-mass searches would remove all triggers within
hundreds or thousands of seconds around each glitch, significantly reducing the
live time of the experiment.

We presented three approximations which allow one to predict the SNR and time
of spurious triggers generated by an inspiral matched filter responding to
sine-Gaussian glitches. Such formulae effectively map the parameters of the
glitch and the chirp mass of the template to the SNR and time of the resulting
spurious trigger. In other words, they represent the first step in understanding
false inspiral triggers produced by isolated, sine-Gaussian glitches. We compared
them to numerical simulations and investigated their validity in the region of
$(f_0, Q, \mathcal{M})$ parameter space relevant for advanced detectors.
Together they complement each other, providing full coverage of the explored
region.

The formulae can be used for vetoing spurious low-mass triggers in the following
way. Assuming we have knowledge of a non-astrophysical, sine-Gaussian-like excitation in the strain
data---either from a burst search pipeline such as Omega \cite{omega}, or by
identifying excitations in auxiliary channels with known couplings to the strain
channel \cite{aux-coupling}---we can predict for each template the SNR and time
of the resulting inspiral triggers. We can then scan the list of triggers
produced by the inspiral search and look for the ones matching the predicted SNR
and time, within a coincidence window accounting for the uncertainty in the
prediction due, for instance, to the detector noise. Triggers found in
coincidence with the prediction can then be associated with the glitch and
selectively removed. Alternatively,
if enough coincident triggers are found, the portion of strain data corrupted by
the glitch can be removed---for instance by replacing it with Gaussian
noise---and the inspiral search repeated. This would essentially remove the
whole cluster of triggers produced by the glitch without sacrificing a segment
of analysis time as long as the longest filter. The more precise definition,
implementation and testing of this procedure constitute another natural followup
of this paper.

\ack

We thank Tom Dent, Drew Keppel, Badri Krishnan, Alex Nielsen and Chris Pankow for
useful discussion and comments. TD is supported by the IMPRS on Gravitational Wave
Astronomy. SVD and SB would like to thank Badri Krishnan and Bruce Allen for
supporting their visits to AEI, Hannover, Germany. SVD would like to thank
IUCAA, Pune, India for a visiting professorship where this work was completed.
This paper has LIGO document number LIGO-P1300016.

\end{document}